\documentclass[11pt]{article}
\usepackage{latexsym}
\usepackage{epsfig}

\def\bg#1{\mbox{\boldmath$#1$}}

\newcommand{\del}{\partial}
\newcommand{\beq}{\begin{eqnarray}}
\newcommand{\eeq}{\end{eqnarray}}
\newcommand{\be}{\begin{eqnarray*}}
\newcommand{\ee}{\end{eqnarray*}}
\newcommand{\bk}{{\bf k}}
\newcommand{\bp}{{\bf p}}
\newcommand{\bq}{{\bf q}}
\newcommand{\br}{{\bf r}}

\newcommand{\ra}{\rightarrow}
\newcommand{\e}{\epsilon}

\newcommand{\nn}{\nonumber}
\newcommand{\ket}[1]{\mbox{$\mid\!#1\rangle$}}
\newcommand{\bra}[1]{\mbox{$\langle#1\!\mid$}}

\begin{document}

\centerline{\Large\bf {Proton-Proton Fusion in Leading Order of}}
\vskip 5mm
\centerline{\Large\bf {Effective Field Theory}}
\vskip 10mm
\centerline{Xinwei Kong$^1$ and Finn Ravndal\footnote{On leave of absence from Institute 
            of Physics, University of Oslo, N-0316 Oslo, Norway}} 
\medskip
\centerline{\it Department of Physics and Institute of Nuclear Theory,}
\centerline{\it University of Washington, Seattle, WA 98195, U.S.A}

\bigskip
\vskip 5mm
{\bf Abstract:} {\small Using a recently developed effective field theory for
the interactions of nucleons at non-relativistic energies, we calculate the rate
for the fusion process $p + p \ra d + e^+ + \nu_e$ to leading order in the momentum 
expansion. Coulomb effects are included non-perturbatively in a systematic way. 
The resulting rate is independent of specific models for the strong interactions 
at short distances and is in agreement with the standard result in the zero-range 
approximation.}

\bigskip
The first step in the different nuclear processes in the Sun which generate the observed
luminosity is proton-proton fusion $p + p \ra d + e^+ + \nu_e$\cite{Bahcall-book}. 
It was explained more than sixty years ago by Bethe and Critchfield\cite{BC} when nuclear 
physics was still in its infancy. When the field had more matured, it was reconsidered in 
the light of more modern developments by Salpeter\cite{ES} and later by Bahcall and 
May\cite{BM}. But in spite of the enormous progress in nuclear physics during this time,
the methods and approximations made in these different calculations were essentially the 
same. The obtained accuracy in the obtained fusion rate was just a few percent. Including 
higher order electromagnetic and strong corrections the uncertainty in the rate is now 
around one percent\cite{BK_1}\cite{Stoks}. This is very impressive for a strongly 
interacting process at low energies very ordinary perturbation theory cannot be used. 

In the light of the importance this fundamental process plays in connection with the solar
neutrino problem and possible neutrino oscillations\cite{Bahcall-book}, it is natural to
reconsider this process from the point of view of modern quantum field theory instead of
the old potential models used previously. A first attempt in this direction was
made by Ivanov et al.\cite{Ivanov}. They obtained then a result which was significantly 
different from the standard result based upon potential models. Subsequently it was pointed 
out by Bahcall and Kamionowski\cite{BK_2} that their effective nuclear interaction was not 
consistent with what is known about proton-proton scattering at low energies where 
Coulomb effects are important.

The approach of Ivanov et al.\cite{Ivanov} is based upon relativistic field theory and 
should in principle yield reliable results. But it is well known that in particular for 
bound states like the deuteron it is very difficult to use consistently a relativistic
formulation. Also the  uncertain nuclear physics part of the fusion process under 
consideration takes place at low energies  and should therefore be described within a 
non-relativistic framework. Then all the large-momentum degrees of freedom are integrated out
and one is left with an effective theory involving only the physically important field 
variables. The underlying, relativistic interactions are then replaced by non-renormalizable 
local interactions with coupling constants which must be determined from experiments 
at low energies. In such a non-relativistic theory the important Coulomb effects can also
systematically be included. 

A step in this direction has recently been taken by Park et al. using chiral perturbation 
theory in the low-energy limit\cite{Park}. They obtain results in very good agreement with
previous potential calculations which is not so surprising since they make use of
phenomenological nucleon wavefunctions which fit low-energy scattering data very well. A
more fundamental approach to the same problem has recently been formulated by Kaplan, 
Savage and Wise in terms of an effective theory for non-relativistic nucleons\cite{KSW_1}. 
It involves a few basic coupling constants which have been determined from nucleon 
scattering data at low energies. With no more free parameters it can then be 
used to make predictions for the deuteron form factor and quadrupole moment\cite{KSW_2}, 
deuteron polarizabilities and Compton scattering on deuterons\cite{gamma}. The obtained 
results are in good agreement with experimental data.
In this approach higher order corrections can also be derived in a systematic way. Going
to higher energies, the effects of pions must be included using the
established counting rules. These will cause the well-known $D$-mixture into the 
deuteron wavefunction.

Most recently, proton-neutron fusion $p + n \ra d + \gamma$ has been calculated in this 
theory by Savage, Scaldeferri and Wise\cite{SSW} including the effects of 
virtual pions. When the process is
taking place at very low energies, one can omit the effects of pions and replace them by 
slightly different couplings of the nucleons alone. From a field-theoretic point of view 
this process is very similar to $p + p \ra d + e^+ + \nu_e$. The main difference is the
strong Coulomb effects which is present in the proton-proton channel. These have now been
calculated and shown to give both a scattering length and an effective range for $pp$ 
low-energy scattering in agreement with data\cite{KR}. Based upon these results, we will
here derive the rate for the corresponding $pp$ fusion reaction. This same process is also 
being considered by Savage and Wise\cite{SW}. At this stage we are only 
interested in the dominant contributions to the fusion rate in order to provide a rough
comparison with results based upon potential models. This corresponds to the zero-range 
approximation in the potential model approach. In light of previous applications, we 
expect the resulting accuracy to be around 20\% or better in the resulting rate. In this
leading order approximation we ignore effective-range corrections, $D$-wave 
admixture, vacuum polarization, two-body current interactions and unknown counterterms.

The strong interactions among the nucleons are now described by the effective Lagrangian
of Kaplan, Savage and Wise\cite{KSW_1}. Denoting the nucleon field of mass $M$ by $N(x)$ 
and including 
only the lowest order interaction term in the $S$-channel, it can be written as
\beq  
     {\cal L} = N^\dagger\left(\del_t + {\nabla^2\over 2M}\right)N
              - C_0(N^T{\bg\Pi}N)\cdot(N^T{\bg\Pi}N)^\dagger               \label{Leff}
\eeq
where the $\Pi_i$  are projection operators into specific spin and isospin states. 
More specifically, for spin-singlet interactions $\Pi_i = \sigma_2\tau_2\tau_i/\sqrt{8}$ 
while for spin-triplet interactions $\Pi_i = \sigma_2\sigma_i\tau_2/\sqrt{8}$.
Calculating now the scattering amplitude for two nucleons, one finds that the coupling
constant $C_0$ is determined by the scattering length $a_{NN}$ in this channel\cite{KSW_1}.
 
For neutron-proton interactions in the spin-triplet channel there is pole in the
corresponding Green's function corresponding to the deuteron. The residue of the
pole gives the bound state wavefunction. With only the lowest order contact interaction
in (\ref{Leff}) it is found to be of the form
\beq
     \psi_d(\bk) = {\sqrt{8\pi\gamma}\over \bk^2 + \gamma^2}                \label{psid}
\eeq
in momentum space. This corresponds to the standard Yukawa form in coordinate space where
$1/\gamma$ represents the size of the deuteron. It determines the value of the corresponding
coupling constant $C_0$ in this channel\cite{KSW_2}.

In the absence of strong interactions, the incoming proton-proton state with center-of-mass
momentum $\bp$ is given by the Coulomb wavefunction\cite{AS} 
\beq
   \psi_\bp(\br) = {1\over\rho}\sum_{\ell = 0}^\infty (2\ell + 1)i^\ell e^{i\sigma_\ell} 
                   F_\ell(\rho)P_\ell(\cos(\theta)                         \label{psip}
\eeq
Here $\rho = pr$ and $\sigma_\ell = \arg{\Gamma(1+\ell + i\eta)}$ is the Coulomb phaseshift 
where the parameter $\eta = \alpha M/2p$ characterizes the strength of the
Coulomb interaction. At low energies only the $S$-wave will contribute. It is given in
terms of the confluent hypergeometric or Kummer function $M(a,b;z)$ as
\beq
    F_0(\rho) = C_\eta \rho e^{-i\rho}M(1-i\eta,2;2i\rho)                   \label{F0}
\eeq
where the normalization factor $C_\eta = e^{-\pi\eta/2}|\Gamma(1 + i\eta)|$. The 
probability $|\psi_\bp(0)|^2$ to find the two protons at the same point is thus equal to 
the Sommerfeld factor\cite{LL} 
\beq
     C_\eta^2 = {2\pi\eta\over e^{2\pi\eta} - 1}                          \label{Sommer}
\eeq
At very low energies when $\eta$ gets large it becomes exponentially small and is the
dominant effect in the fusion reaction.
    
The available energy in the process is set by the neutron-proton mass difference
and the deuteron binding energy $B$ = 2.225 MeV. It corresponds to a momentum $\gamma = 
\sqrt{BM}$ of the bound nucleons equal to 45.71 MeV. The temperature in the core of the Sun is 
approximately $15\times 10^6$\,K which corresponds to an average proton momentum around 
$p = 1.5$\, MeV. The kinetic energy of the lepton pair will therefore be much smaller
than $\gamma$ and it is a very good approximation to just ignore it. 
In the following the weak current is therefore assumed to carry zero momentum
and the evaluation of the different Feynman diagrams will much simplify.

To lowest order in the strong interaction between the protons, the transition matrix $T_{fi}$
is given by the first diagram in Fig.1. After being hit by the weak current,
the proton-proton system is transformed into a bound deuteron. The value of the diagram is
thus seen to be
\beq
      A(p)  = \sqrt{8\pi\gamma}\int\!{d^3k\over(2\pi)^3} 
           {1\over \bk^2 + \gamma^2}\psi_\bp(\bk)                    \label{TA}
\eeq
where $\psi_\bp(\bk)$ is the Fourier transformed wavefunction of the incoming 
proton-proton scattering state. To first order in the four-proton $^1S_1$ coupling 
$C_0$ we have the diagram in Fig. 1b.
\begin{figure}[htb]
 \begin{center}
  \epsfig{figure=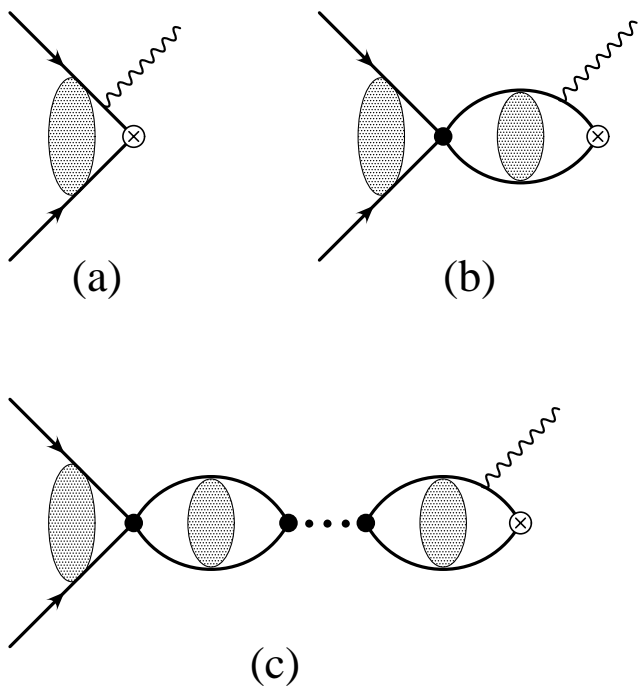,height=90mm}
 \end{center}
 \vspace{-5mm}
 \caption{\small Feynman diagrams which contribute in leading order to proton-proton
fusion. The solid lines are nucleons and the wiggly line represents the weak current.
Between the protons there are exchanged Coulomb photons represented by the shaded blobs
while the crossed circles represents the deuteron wavefunctions.}
 \label{fig1}
\end{figure}
Within the loop the protons move in the Coulomb field of each other. This motion is
described by the standard Coulomb propagator $G_C(E) = 1/(E - H_0 - V_C + i\e)$ where
$H_0$ is the free, non-relativistic Hamiltonian and $V_C$ is the Coulomb potential. In
momentum space it takes the form
\beq
     G_C(E;\bk,\bk') =  M\!\int\!{d^3 q\over (2\pi)^3}
                       {\psi_\bq(\bk)\psi_\bq^*(\bk')\over\bp^2 -\bq^2 + i\e}  \label{GC}
\eeq
when expressed in terms of the Coulomb wavefunctions for the center-of-mass 
energy $E= p^2/M$. As illustrated in Fig.2 it includes the free propagator plus the effects 
\begin{figure}[htb]
 \begin{center}
  \epsfig{figure=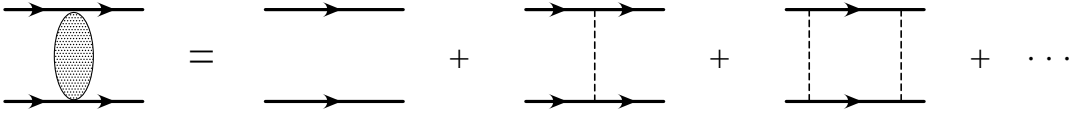,height=15mm}
 \end{center}
 \vspace{-5mm}
 \caption{\small The full Coulomb propagator is formed from the infinite sum
of exchanged static photons.}
 \label{fig2}
\end{figure}
of one, two and more exchanged static photons. In this way we find for the diagram in Fig.1b
the value $C_0B(p)$ where
\beq
    B(p)  =  \sqrt{8\pi\gamma}\int\!{d^3k\over(2\pi)^3}{d^3k'\over(2\pi)^3} 
           {1\over \bk^2 + \gamma^2}G_C(E;\bk,\bk')                \label{TB}
\eeq
is a convergent integral. 

The protons can suffer another rescattering as in Fig.1c before they are converted to
a deuteron by the weak interaction. Denoting this extra bubble in the diagram by $J_0(p)$,
the full diagram is then just $C_0J_0$ times the previous diagram. The magnitude of 
the bubble equals the 
probability amplitude for the two protons to propagate from zero spatial separation back to
zero separation, i.e. $J_0(p) = G_C(E;0,0)$. In terms of the representation (\ref{GC}) of 
the Coulomb propagator it is given by the integral
\beq
     J_0(p) = M\!\int\!{d^3 q\over (2\pi)^3} {2\pi\eta(q)\over e^{2\pi\eta(q)} - 1}
               {1\over p^2 - q^2 + i\e}                                    \label{J0}
\eeq
which has an ultraviolet divergence. After regularization, it will give a renormalization
of the coupling constant $C_0$.

We can now continue to add in more such rescattering diagrams in Fig.1c. They form a geometric
series which sums up to $C_0/(1 - C_0J_0)$. All the diagrams including the 
first then gives for the hadronic part of the full transition amplitude
\beq
    T_{fi}(p) =  A(p) + B(p){C_0\over 1 - C_0J_0(p)} \psi_\bp(0)     \label{Tfi}
\eeq
where the last factor  $\psi_\bp(0) = C_\eta e^{i\sigma_0}$ gives the
amplitude for the two incoming protons to meet at the first vertex.
It can be written in a more compact and 
recognizable form by introducing the proton-proton scattering state 
\beq
    \ket{\Psi_\bp} = [1 + \sum_{n=1}^\infty(G_CV_0)^n ]\ket{\psi_\bp}
\eeq
where the strong interaction potential $\bra{\bp}V_0\ket{\bq} = C_0$ is included to all
orders in addition to the Coulomb interaction. Then we see that the matrix element 
(\ref{Tfi}) is just the overlap integral between this wavefunction and the deuteron
wavefunction (\ref{psid}), 
\beq
     T_{fi}(p) = \int\!{d^3 k\over (2\pi)^3}\psi_d^*(\bk)\Psi_\bp(\bk)       \label{overlap}  
\eeq
as follows from the expressions in (\ref{TA}) and (\ref{TB}). 
This form of the transition 
matrix element was written down first by Bethe and Chritchfield\cite{BC} and used subsequently 
by everyone considering the process in potential models.

We can now evaluate the different parts of the transition matrix element (\ref{Tfi}). The first
part (\ref{TA}) is most easily found in coordinate space where we have the Coulomb wavefunction
(\ref{F0}). It gives
\beq
     A(p) &=&  \sqrt{8\pi\gamma}C_\eta e^{i\sigma_0}\int_0^\infty\!dr r e^{-(\gamma +ip)r}
             M(1-i\eta,2;2ipr) \\ \nn
          &=&   {\sqrt{8\pi\gamma}C_\eta e^{i\sigma_0}\over (\gamma + ip)^{2}}
                 {_2F_1}\left(1-i\eta,2;2;{2ip\over\gamma + ip}\right)
\eeq
Now the hypergeometric function $_2F_1(a,b,b;z) = (1 - z)^{-a}$ so that the final result 
can be written as
\beq
    A(p) = C_{\eta}e^{i\sigma_0}{\sqrt{8\pi\gamma}\over p^2 +\gamma^2}
            e^{2\eta\arctan({p\over\gamma})}                                \label{Ap}
\eeq
In the expression (\ref{TB}) for $B(p)$ we notice that the integral over $\bk'$ gives 
the complex conjugate value of the Coulomb wavefunction at the origin. It therefore takes 
the form
\be
    B(p) =   M \int\!{d^3k\over(2\pi)^3}\int\!{d^3q\over(2\pi)^3} 
            {\sqrt{8\pi\gamma}\over \bk^2 + \gamma^2}{\psi_\bq(\bk)\over\bp^2 -\bq^2 + i\e} 
           \psi_\bq^*(0)     
\ee
The integral over $\bk$ is just the previous result for $A(q)$ so that
\beq
    B(p) = M \int\!{d^3q\over(2\pi)^3}{\sqrt{8\pi\gamma}\over q^2 +\gamma^2} 
            {e^{2\eta\arctan({q\over\gamma})}\over p^2 - q^2 + i\e}
            {2\pi\eta(q)\over e^{2\pi\eta(q)} - 1}                            \label{Bp}
\eeq
It is seen that when the momentum of the incoming protons is non-zero the integral
yields a complex result.

The infinite sum over proton bubble diagrams in Fig.1 is just the proton-proton strong
scattering amplitude modified by Coulomb corrections\cite{KR}. It is given by the
bubble integral (\ref{J0}). In order to regularize it we use the special PDS scheme constructed
by Kaplan, Savage and Wise for this effective theory\cite{KSW_1}. It is based on ordinary 
dimensional regularization around $d=3$ dimensions. The difference lies in that poles in 
$d=2$ dimensions are subtracted. This gives rise to terms which depend on the regularization
point $\mu$. In the present case we obtain with $\e = 3-d$
\beq
    J_0(p) = {\alpha M^2\over 4\pi}\left[{1\over\e} + \ln{\mu\sqrt{\pi}\over\alpha M} + 1
             - {3\over 2}C_E - H(\eta) \right] - {\mu M\over 4\pi}              \label{J01}
\eeq
Here $C_E = 0.5772\ldots$ is Euler's constant and the function
\beq
    H(\eta) = \psi(i\eta) + {1\over 2i\eta} -\ln(i\eta)                         \label{Heta}
\eeq
The divergent $1/\e$ piece will be absorbed in counterterms representing electromagnetic
interactions at shorter scales. This replaces the bare coupling constant $C_0$ with the
renormalized value $C_0(\mu)$. Matching the calculated proton-proton scattering amplitude 
to the experimental one, we can determine this coupling constant in terms of the measured
scattering length $a_p$ which gives the cross-section in the zero-energy limit\cite{KR},
\beq
    {4\pi\over MC_0(\mu)} = {1\over a_p} - \mu  
    + \alpha M\left[\ln{\mu\sqrt{\pi}\over\alpha M} + 1 - {3\over 2}C_E \right]
\eeq
The part of the scattering amplitude which is needed in (\ref{Tfi}) is therefore
\beq
    C_0^{-1}(\mu) - J_0(p) = {M\over 4\pi}\left[{1\over a_p} + \alpha M H(\eta)\right]
\eeq
This is in general complex because of the function (\ref{Heta}). Its real part is the more
phenomenologically relevant function $h(\eta) = \mbox{Re}\,\psi(i\eta) - \ln \eta$ while 
the imaginary part is simply $C_\eta^2/2\eta$. At very low 
energies the real part $h(\eta) = 1/(12\eta^2) + {\cal O}(\eta^{-4})$ dominates since the
imaginary part is then exponentially small. These functions are well-known in the 
context of proton-proton scattering\cite{JB}. 

For the very small proton momenta we have in the Sun, one evaluates the transition matrix
element just as well at zero momentum\cite{BM}. The first term (\ref{Ap}) is then simply
\beq
    | A(p\ra 0) | = \sqrt{\frac{8\pi C_{\eta}^2}{\gamma^3}} \, e^\chi
\eeq
where the parameter $\chi =\alpha M/\gamma$. It is therefore natural to introduce 
the standard reduced matrix element\cite{ES}
\beq
    \Lambda(p) = \sqrt{{\gamma^3\over 8\pi C_{\eta}^2}} \, |T_{fi}(p)|
\eeq
Using $2 \pi \eta(q)$ as a new integration variable in the expression (\ref{Bp}) for $B(p\ra 0)$,  
it becomes proportional to the integral
\beq
    I(\chi) = \int_0^\infty\!dx {2x\over e^x -1}
    {e^{{x\over\pi}\arctan({\pi\chi\over x})}\over x^2 + \pi^2\chi^2}      \label{Int}
\eeq
which we can only do numerically. We thus have the result
\beq
    \Lambda(0)= e^\chi - \alpha M a_p\,I(\chi)                            \label{Lambda0}
\eeq
The first part is identical with what one obtains in potential models\cite{BM} 
while the second part seems to have a different dependence on the parameter $\chi$. However,
by numerical integration, we find that it is in fact exactly the same. So far we have
been unable to show this equality by analytical methods.

However, it can be demonstrated by taking first the zero energy limit of the proton-proton
wavefunction in the integrand of the transition matrix element (\ref{overlap}) and then 
afterwards integrate. The regular Coulomb wavefunction (\ref{F0}) 
simplifies then to\cite{AS}
\beq
     F_0(\rho) \simeq {C_\eta\over \sqrt{2\eta}}\rho^{1/2}I_1(2\sqrt{2\eta\rho})
\eeq
when $2\eta \gg \rho$. The first part (\ref{TA}) of the matrix element will now be given
by the integral
\be
   A(p\ra 0) = C_\eta e^{i\sigma_0} \sqrt\frac{4\pi\gamma}{\eta}{1\over p^2}
\int_0^\infty\!d\rho
               \rho^{1/2}e^{-\gamma\rho/p}I_1(2\sqrt{2\eta\rho})
\ee
which can be expressed in terms of a Whittaker function,
\be
   |A(p\ra 0)| =  \sqrt{\frac{8\pi C_{\eta}^2}{\gamma^3}}  
\frac{e^{\chi/2}}{\chi}  M_{-1,{1\over 2}}(\chi)
\ee
Now $M_{-1,{1\over 2}}(\chi) = \chi e^{\chi/2}$ and thus we reproduce the first part of
(\ref{Lambda0}).

In order to calculate $B(p\ra 0)$ we go back to the result (\ref{TB}). Now we use a different
representation of the Coulomb propagator constructed from the regular $F_\ell(\rho)$ and the 
irregular $G_\ell(\rho)$ eigenfunctions\cite{Brown}. We only need the radial part in the 
$S$-channel 
\beq
     G_C(E;r,r')_{\ell = 0} = - {Mp\over 4\pi}{F_0(\rho_<)\over \rho_<}{G_0(\rho_>)\over \rho_>} 
\eeq
where $r_<$ and  $r_>$ are the smallest and largest of $r$ and $r'$ respectively.
The irregular solution $G_\ell(\rho)$ is normalized so that the Wronskian 
$G_0F_0^{'} - F_0G_0^{'}= 1$. In (\ref{TB}) this Green's function is seen to enter with 
the argument $r' = 0$. Since $F_0(\rho)/\rho = C_\eta$ in the limit $\rho \ra 0$, it then 
simplifies to
\be
    G_C(E;r,0)_{\ell = 0} = - C_\eta {MpG_0(\rho)\over 4\pi\rho}
\ee
In the zero-energy limit $2\eta \gg \rho$ we can then use
\beq
     G_0(\rho) \simeq {2\over C_\eta}\sqrt{2\eta\rho}K_1(2\sqrt{2\eta\rho})
\eeq
For the function $B(p)$ we thus find in this limit
\be
    B(p\ra 0) = - {M\over p}\sqrt{4\gamma\eta\over\pi}
   \int_0^\infty\!d\rho \rho^{1/2}e^{-\gamma\rho/p}K_1(2\sqrt{2\eta\rho})
\ee
This integral is now given by the other Whittaker function,
\be
    B(p\ra 0) = - \frac{M^2 \alpha}{4 \pi}  \sqrt{\frac{8 \pi}{\gamma^3}} 
 \frac{e^{\chi/2}}{\chi} W_{-1,{1\over 2}}(\chi)
\ee
Since $W_{-1,{1\over 2}}(\chi) = W_{-1,-{1\over 2}}(\chi)$ we can thus express the result 
by the simpler integral
\beq
     W_{-1,-{1\over 2}}(\chi) = \chi e^{\chi/2}\int_\chi^\infty\!dt {e^{-t}\over t^2}
\eeq
By partial integration it can be written in terms of the exponential integral
function $E_1(\chi)$. In this way we finally obtain for the original integral (\ref{Int}) 
\beq
     I(\chi)= {1\over\chi} - e^\chi E_1(\chi)
\eeq
in agreement with the standard result\cite{BM}.
  
With the previous value $\gamma = 45.71\,$MeV for the deuteron, we have
$\chi = 0.15$ and the integral $I(0.15)= 4.96$. Combined with the measured 
value $a_p = -7.82$\,fm for the scattering length we then have $\Lambda(0) = 2.51$ for the
reduced matrix element. A recent and most accurate calculation including higher order
corrections based on a potential model\cite{Stoks} gives a value 
$\Lambda^2(0) = 7.05 \pm 0.02$. This corresponds to $\Lambda(0) = 2.66$. Our leading order 
result from effective field theory which we have found to agree with the zero-range 
approximation of potential models, is thus accurate to within 6\%  percent of the full 
result.

In next order of the momentum expansion of the underlying effective field theory it is 
not clear if the transition matrix element can still be given as a simple
overlap integral of the two wavefunctions as in (\ref{overlap}).  The corresponding Feynman 
diagrams for the transition matrix element are also technically more difficult to evaluate 
because they involve the Coulomb propagator in new ways. But it is important to calculate
these corrections and verify that they are as small as is found in potential models.

We want to thank John Bahcall, Jiunn-Wei Chen, Harald Griesshammer, Daniel Phillips, 
Martin Savage and Mark Wise for many helpful discussions and comments. In addition, 
we want to thank the Department of Physics and the INT for generous support and hospitality. 
Xinwei Kong is supported by the Research Council of Norway.

\end{document}